\documentclass{apex}
\title{Multi-Junction Switching in Bi$_2$Sr$_{1.6}$La$_{0.4}$CuO$_{6+\delta}$ Intrinsic Josephson Junctions}
\author{\textsc{Hiromi Kashiwaya},
\textsc{Tetsuro Matsumoto},
\textsc{Hajime Shibata},
\textsc{Hiroshi Eisaki},
\textsc{Yoshiyuki Yoshida},
\textsc{Hiroshi Kambara},
\textsc{Shiro Kawabata}$^{1}$, and 
\textsc{Satoshi Kashiwaya}}
\inst{Nanoelectronics Research Institute (NeRI), National Institute of Advanced Industrial Science and Technology (AIST), Central 2, Tsukuba, Ibaraki 
305-8568, Japan\\
$^{1}$Nanotechnology Research Institute, AIST, Central 2, Tsukuba, Ibaraki 305-8568, Japan\\
}
\abst{
We study the dynamics of multi-junction switching (MJS): several intrinsic Josephson junctions (IJJs) in an array switch to the finite voltage state simultaneously.
The number of multi-switching junctions ($N$) was successfully tuned by changing the load resistance serially connected to an Bi$_2$Sr$_{1.6}$La$_{0.4}$CuO$_{6+\delta}$ IJJ array.
The independence of the escape rates of $N$ in the macroscopic quantum tunneling regime indicates that MJS is a $successive$ switching process rather than a $collective$ process.
The origin of MJS is explained by the gradient of a load curve and the relative magnitudes of the switching currents of quasiparticle branches in the current-voltage plane.
}

\begin{document}
%
\maketitle 
%
%
%
%
%
Much work is focused on the switching dynamics of current-biased Josephson junctions (JJs) made of cuprate superconductors.
An intrinsic Josephson junction (IJJ) that utilizes the $c$-axis transport of cuprates\cite{Kleiner} is the most promising candidate for realizing a high-$T_c$ qubit\cite{Inomata,Bauch}.
It is well known that the advantages of the IJJ are its high critical temperature and the excellent quality of the insulator.
On the other hand, the appearance of distinctive switching dynamics is expected owing to the coupling effect between adjacent IJJs in an array, because IJJs are separated by only a few atomic layers.
Therefore clarifying the switching dynamics in IJJs under the influence of the coupling effect is an important issue for realizing a high-$T_c$ qubit.
Reflecting the configuration, the current-voltage ($I$-$V$) curve of an IJJ array exhibits a multiple-branch structure [see Fig. 2(a)].
Here we denote the critical current and switching current of the $j$th branch as $I_{cj}$ and $I_{swj}$, respectively ($j$=0,1,2,..).
\par
%
%
Two types of distinctive dynamics have been reported for Bi$_2$Sr$_{2}$CaCu$_{2}$O$_{8+\delta}$ (Bi2212) IJJs thus far.
One is a $successive$ switching process, in which each single junction switching (SJS) occurs at different times when the bias current is gradually increased.
This process has been observed in IJJs of an array having nonuniform critical currents ($I_c$s).
The properties of the first switching (0th branch to 1st branch) are mostly consistent with those of conventional JJ switching: switching due to the thermal activation above the crossover temperature ($T^*$) and due to the macroscopic quantum tunneling (MQT) below $T^*$\cite{Inomata,Bauch,hiromiJPSJ}.
However, for higher-order switching, that is, the switching from the $j$th branch ($j\ge$1), the switching properties are significantly modified by the effect of the IJJs in the finite voltage state\cite{hiromiJPSJ,OtaPRB,warb}.
A large enhancement of the fluctuation, detected as an increase in the escape temperature, seriously suppresses the switching current ($I_{swj}$, $j\ge$1 ) as compared to the conventional JJ switching cases.
The other distinctive dynamics is a $collective$ switching process, referred to as uniform junction switching, in which all IJJs in an array switch simultaneously without staying in any intermediate stable states.
This process has been observed in IJJs with uniform values of $I_c$\cite{Jin}.
The escape rate of the $collective$ switching shows $N^{2}$ times as large as that of SJS in the quantum regime. 
Here, $N$ represents the number of junctions switching simultaneously.
The significant enhancement of the escape rate indicates the presence of peculiar tunneling processes\cite{Fistul,Savelev,Koyama}.
%
Also the mechanism why the intermediate stable states are skipped in this uniform junction switching is still unclear.
%
%
%
\begin{figure}
\includegraphics[width=0.8\linewidth]{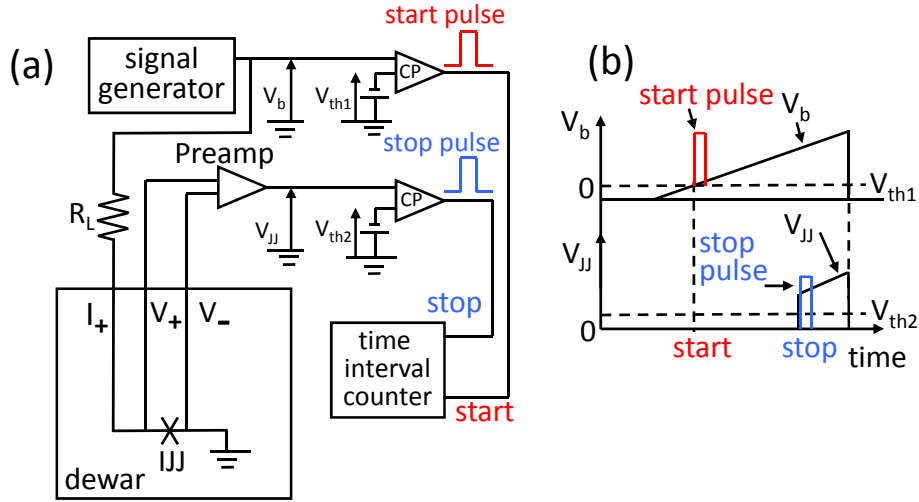}
\caption{\label{fig:setup} 
(a) Schematic of the electric circuit used in the measurements.
(b) Schematic of the time sequence illustrating the bias voltage ($V_{b}$) and the junction voltage ($V_{JJ}$). 
The start pulse is generated when $V_{b}$ exceeds 0 mV. 
The stop pulse is generated when the junction voltage exceeds $V_{th2}$.
}
\end{figure}
\par
In this paper, we report on a novel type of switching process, in which the simultaneous switching of not all but only several of the IJJs in an array occurs.
We refer to this process as multi-junction switching (MJS).
We found that the number of simultaneously switching junctions $N$ in MJS can be tuned by changing the load resistance ($R_{L}$) serially connected to the IJJs in the measurement circuit.
The switching current distribution and the escape rate of MJS are systematically measured as functions of $N$ to reveal the differences among various switching processes.
\par
%
%
\begin{figure}[t]
\includegraphics[width=0.8\linewidth]{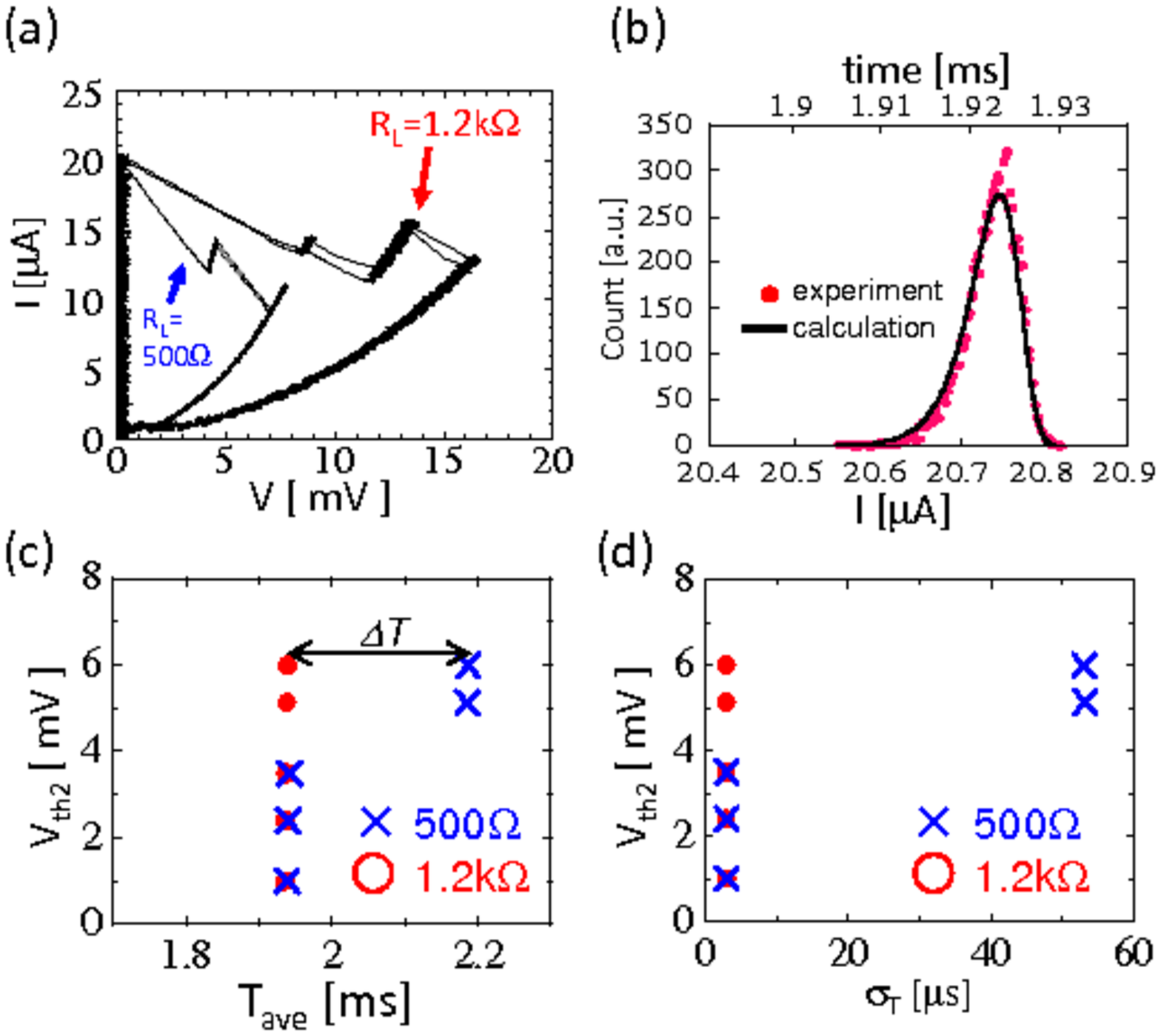}
\caption{\label{fig:howtoN} 
(a) Typical $I$-$V$ curves of the Bi2201 IJJ in the cases of $R_{L}$=500 $\Omega$ and 1.2 k$\Omega$ measured at $T$=40 mK. 
(b) Typical switching current (or time interval) distribution obtained at T=40 mK. Points represent experimental data. Solid lines represent the theoretical fitting at MQT region.
(c) (d) $R_{L}$ dependences of the average time ($T_{ave}$) and the standard deviation of time ($\sigma_{T}$) as functions of $V_{th2}$ obtained from the time interval distribution at $T$=40 mK.
}
\end{figure}
%
%
%
%
%
%
\par
We used Bi$_2$Sr$_{1.6}$La$_{0.4}$CuO$_{6+\delta}$ (Bi2201) IJJs for the present study because both the nonequilibrium effect and the coupling strength due to the electromagnetic effect are enhanced for single-layered superconductors.
The Bi2201 IJJs were fabricated by a focused ion beam (FIB) etching process from a single crystal grown by the floating zone method.
Details of the fabrication process of Bi2201 IJJs are described in ref. \citen{tetsuAPEX}.
A schematic of the electric circuit is shown in Fig. 1, and details of the measurement system are described in ref. \citen{hiromiIEICE}. 
The bias current was introduced from the voltage source to the junction via a load resistance of $R_{L}$.
A trigger circuit generated a start pulse when the bias voltage ($V_{b}$), which was ramped up at a constant rate, exceeded $V_{th1}$. 
The amplified junction voltage ($V_{JJ}$) was forwarded to a comparator (CP), which generated a stop pulse when the output signal exceeded $V_{th2}$ as shown in Fig. 1(b).
The time interval between the start and stop pulses was measured by a time-interval counter.
The time distribution can be converted to a switching current distribution\cite{Wal, hiromiJPSJ} shown as Fig. 2(b).
Typical sweep rate was 0.011A/sec and typical number of samplings to obtain the distribution was 10000.
For the measurements of higher-order switching processes, $V_{th2}$ was tuned to an appropriate value.
\par
%
Figure 2(a) shows typical $I$-$V$ curves of a Bi2201 IJJs for $R_{L}$=500 $\Omega$ and 1.2 k$\Omega$ obtained at $T$=40 mK.
It is important to note that two curves are obtained for the same sample.
The $I$-$V$ curves appear to correspond to $N$=1 for $R_{L}$=500 $\Omega$ and $N$=2 for 1.2 k$\Omega$.     
To remove the ambiguity of $N$, we measured the average time ($T_{ave}$) and standard deviation ($\sigma_{T}$) of the time interval distribution as functions of $V_{th2}$.
$T_{ave}$ and $\sigma_{T}$ obtained for different $R_{L}$ values are shown in Figs. 2(c) and 2(d).
According to our previous study on SJS\cite{hiromiJPSJ}, a large enhancement of $\sigma_{T}$ is expected for higher-order switching processes because of the enhanced fluctuation.
Actually, when $R_{L}$=500 $\Omega$, $T_{ave}$ has a time lag ($\Delta T$) and $\sigma_{T}$ is enhanced when $V_{th2}$ exceeds 4 mV as shown in Figs. 2(c) and 2(d).
This result indicates that the first switching (0th branch to 1st branch) and the second switching (1st branch to 2nd branch) occur separately and that there is an intermediate stable state on the first branch [see Fig. 3(c), LC$_1$].
Note that $\Delta T$ of Fig. 2(c) also gives information of the actual time staying in the quasiparticle branch.
In contrast, when $R_{L}$=1.2 k$\Omega$, neither $T_{ave}$ nor $\sigma_{T}$ change, even for $V_{th2}>$4 mV, within an accuracy of 1$\times$10$^{-7}$ s.
The simultaneous switching without staying intermediate states is a manifestation of MJS corresponding to switching from 0th branch to 2nd branch [see Fig. 3(c), LC$_2$].
It is interesting that the number of simultaneously switching junctions can be tuned simply by changing the magnitude of $R_{L}$.
%
%
%
%
\begin{figure}[t]
\includegraphics[width=0.8\linewidth]{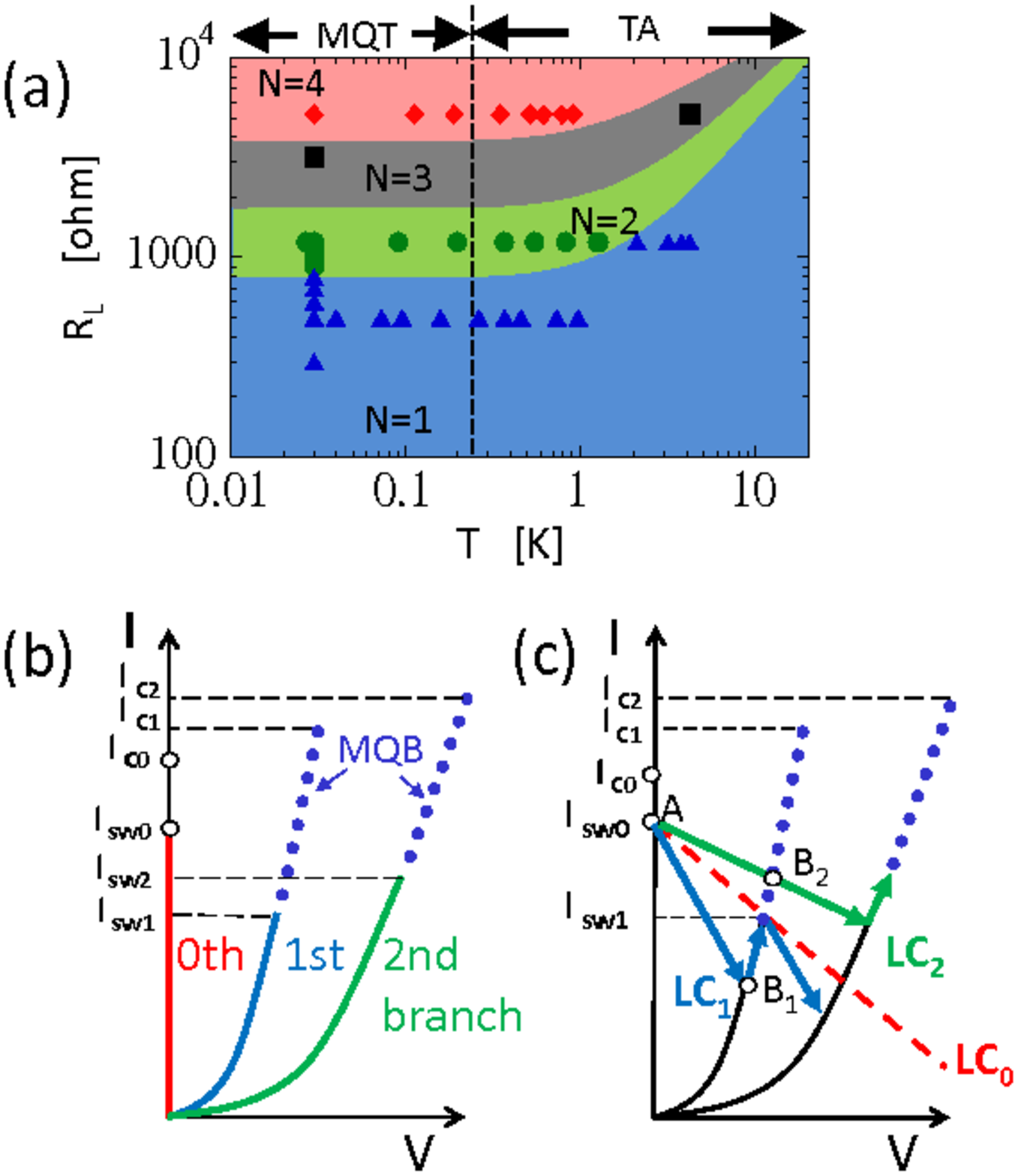}
\caption{\label{fig:Ndepend}
(a) Schematic diagram of $N$ as a function of $R_{L}$ and temperature. 
Symbols in the figure correspond to experimentally detected points.
(b) Schematic of the $I$-$V$ curve and the relative magnitude of the critical currents and the switching currents.
Missing quasiparticle branches (MQBs) corresponding to the unstable states ($I>I_{sw1}$,$I_{sw2}$) are shown as dotted lines.
(C) Schematic illustration of the SJS and the MJS processes in the $I$-$V$ plane. 
The trajectories of SJS and MJS while ramping up the bias voltage are LC$_1$ and LC$_2$, respectively.
As the bias voltage is gradually increased, the state jumps from A on the 0th branch to B$_1$ or B$_2$ on the first branch along the load curve whose gradient is inversely proportional to $R_L$.
For the case of LC$_1$, SJS occurs because B$_1$ is stable, while for the case of LC$_2$, MJS occurs because B$_2$ is on the MQB.
}
\end{figure}
\par
Figure 3(a) shows a schematic diagram of $N$ as a function of $R_{L}$ and the temperature determined by the above-mentioned procedure. 
%
Symbols in the figure correspond to experimentally detected points.
It is clear that $N$ increases with increasing $R_{L}$ for a fixed temperature.
Taking account of this feature, the origin of the MJS process can be phenomenologically explained as follows.
%
The most important point is the relative magnitudes of the $I_c$s and $I_{sw}$s, as schematically shown in Fig. 3(b).
%
$I_{c0}$ is always smaller than $I_{c1}$ and $I_{c2}$ ($I_{c0}<I_{c1}, I_{c2}$) because each junction switches to the finite voltage state in order of increasing $I_c$ when the bias voltage is ramped up.
%
In contrast, since the escape temperature for higher-order switching processes is markedly increased, $I_{sw1}$ and $I_{sw2}$ are considerably reduced from $I_{c1}$ and $I_{c2}$, respectively\cite{hiromiJPSJ}.
%
This enhanced escape temperature can induce the reversal of the orders of the magnitudes, giving $I_{sw0}>I_{sw1}$, $I_{sw2}$.
%
The states on the first and second branches are no longer stable for $I_{c1}>I>I_{sw1}$ and $I_{c2}>I>I_{sw2}$, respectively.
%
The MJS process is promoted by these `` missing quasiparticle branches (MQBs)".
%
Figure 3(c) schematically illustrates the trajectories of the SJS and the MJS processes while ramping up the bias voltage.
%
When the bias voltage is gradually increased, the junction state jumps from A on the 0th branch to B$_1$ or B$_2$ on the 1st branch along the load curve whose gradient is inversely proportional to $R_L$.
The line connecting $I_{sw0}$ and $I_{sw1}$, labeled as LC$_0$ (693.2 $\Omega$ in the present IJJ) in Fig. 3(c), corresponds to the critical load curve that separates SJS and MJS.
When $R_{L}<R_{L0}$, the junction state jumps from A to B$_1$ on the stable quasiparticle branch; therefore MJS does not occur.
%
In contrast, when $R_{L}$ is considerably larger than $R_{L0}$, the state jumps from A to B$_2$ that locates inside the MQB, then the next jump immediately occurs without staying on the branch.
This immediate switching is the origin of MJS.
Actually, the critical load curve obtained in this scheme is comparable to experimentally obtained value.
Meanwhile, when $R_{L}$ is comparable to $R_{L0}$, more complex dynamics is expected, which will be described in a separate paper.
\par
%
%
%
\begin{figure}[]
\includegraphics[width=0.8\linewidth]{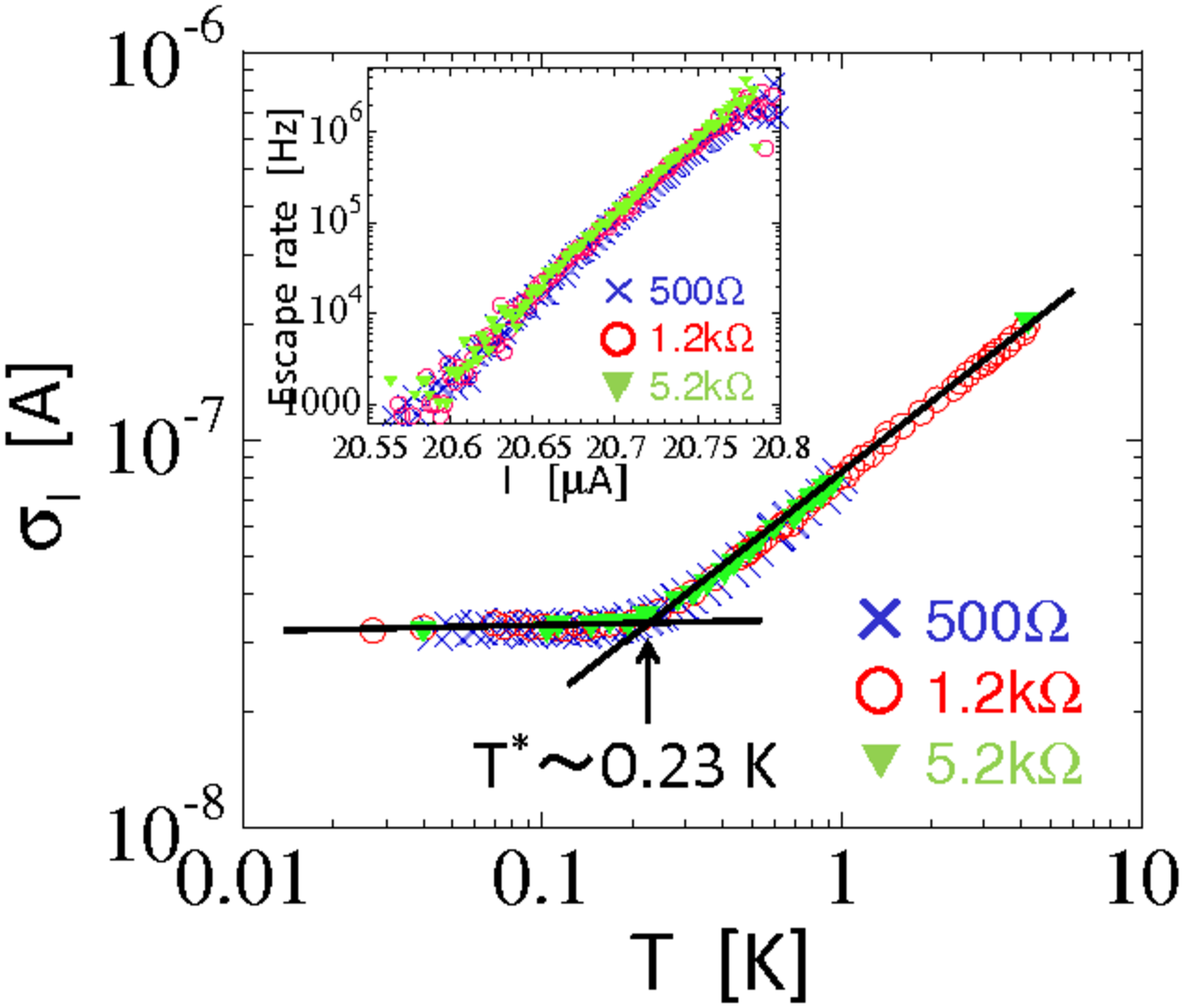}
\caption{\label{fig:IsigmaT} 
Temperature dependence of $\sigma_{I}$ defined by the standard deviation of the switching probability distribution $P(I)$ for $R_{L}$ = 500 $\Omega$ (crosses), 1.2 k$\Omega$ (circles), and 5.2 k$\Omega$ (triangles).
$T^{2/3}$ dependence in the thermal activation regime and saturation in the MQT regime (below 0.23 K) are detected.
The inset shows the escape rates for $N$=1, 2 and 4 at 40 mK.
}
\end{figure}
%
%
\par
The above analysis suggests that the MJS process originates from the stability of higher-order branches rather than the onset of the collective switching process.
To confirm this scheme, we measured the temperature dependence of the standard deviation of the switching current distribution $P$($I$) for $R_{L}$=500 $\Omega$, 1.2 k$\Omega$ and 5.2 k$\Omega$ as shown in Fig. 4. 
%
$V_{th2}$ was fixed at 1 mV.
%
For all cases, the switching current distribution forms a single curve that has $T^{2/3}$ dependence above $T^* \sim$0.23 K dominated by the thermal activation and is saturated below $T^*$ dominated by the MQT.
%
Theoretically calculated $T^*$ of 0.29 K based on eq. 19 of ref. \citen{Wal} using the plasma frequency of 88GHz, the capacitance of 212 fF, and quality factor of 22 obtained by microwave irradiation is consistent with the experimental value.
%
These results confirm that the switching features of MJS mostly agree with those of conventional JJs. 
The $N^{2}$ enhancement of the escape rate expected for $collective$ switching cannot be detected at present.
%
Therefore, we conclude that each switching in the MJS process takes place $successively$ rather than $collectively$. 
However, MJS is distinct from conventional $successive$ switching in the point that the enhancements of the fluctuation in the higher order switching processes are completely absent as previously mentioned.
\par
%
An important question is why the $collective$ switching process reported in ref. \citen{Jin} cannot be observed in the present measurements.
%
We speculate that this inconsistency stems from the difference in the uniformity of the values of $I_c$ in IJJs in an array.
%
The values of $I_c$ in an IJJ array fabricated by the FIB process tend to be more widely distributed than those fabricated by Ar ion milling\cite{Jin}.
%
Actually, the estimated values of $I_{c0}$ and $I_{c1}$ based on the Kramers escape rate and the MQT models are 21.2 and 21.5 $\mu$A, respectively.
The present experimental results verify that these values are sufficiently close for the emergence of MJS, but probably not sufficiently close for the emergence of the $collective$ switching process.
Therefore, it will be interesting to clarify how much uniformity is required to observe the $collective$ process.
This issue will be confirmed in future works.
%
%
\par

\acknowledgments
This work was financially supported by a Grant-in-Aid for Young Scientists (b)(No. 21710100) from Japan Society for the Promotion of Science, Japan, and by Mitsubishi foundation.

\end{document}